\input{aipcheck}

\documentclass[
    ,final            
  ]
  {aipproc}

\layoutstyle{6x9}

\def\be{\begin{eqnarray}&&}
\def\ee{\end{eqnarray}}
\def \nonu {\nonumber \\&&}

\begin{document}

\title{Trinucleon Electromagnetic  Form Factors  
 and the Light-Front Hamiltonian Dynamics }


\author{F. Baroncini}{
  address={Dip. di Fisica, Universit\`a di Roma "Tor Vergata" , Via della Ricerca
Scientifica 1, I-00133  Roma, Italy}
}
\author{A. Kievsky}{
  address={Istituto Nazionale di Fisica Nucleare, Sezione di Pisa,Via Buonarroti
  2, I-56100 Pisa, Italy}
}

\author{E. Pace}{
  address={Dip. di Fisica, Universit\`a di Roma "Tor Vergata" \& INFN - Sezione di Tor
  Vergata, Via della Ricerca
Scientifica 1, I-00133  Roma, Italy}
}
\author{G. Salm\`e}{
  address={Istituto Nazionale di Fisica Nucleare, Sezione di Roma, P.le A. Moro 2,
 I-00185 Roma, Italy } 
}

\begin{abstract}
 This contribution briefly illustrates  preliminary calculations of 
 the electromagnetic form factors of  $^3$He and 
  $^3$H, obtained  within
  the Light-front Relativistic Hamiltonian Dynamics,
  adopting i) a Poincar\'e
 covariant current operator, without dynamical two-body currents, and ii)
    realistic
 nuclear bound states with $S$, $P$ and $D$ waves. The kinematical region of 
 few $(GeV/c)^2$, 
 relevant for 
 forthcoming TJLAB experiments, has been investigated, obtaining possible
 signatures of relativistic effects for $Q^2>2.5 ~(GeV/c)^2$.
\end{abstract}

\maketitle


\section{Introduction}

Relativistic Hamiltonian
Dynamics (RHD), introduced  by Dirac in a seminal paper \cite{Dirac},
represent a viable tool for fulfilling the Poincar\'e covariance, i.e. a
fundamental  symmetry. In particular, in the study of the 
electromagnetic (em)
form factors  (ff) of few-nucleon systems some advancements were made within
the so-called Light-front (LF) RHD (one out the three ones  proposed by Dirac), 
 both calculating
the Deuteron em observables \cite{LPS2K} and obtaining a first description of the
Trinucleon ff \cite{BPS07} (retaining only $S$ and $S^\prime$ waves). 
It should be pointed out that relativistic calculations have a twofold interest,
 both from a general point of view and from a phenomenological one,
   given the forthcoming experiments at TJLAB, in the region of few $(GeV/c)^2$  
    (see, e.g.,  Ref.  \cite{Makis} for the $^3$He and 
  $^3$H ff).
   
 Our aim is  to construct, within the LF RHD, a relativistic 
 approach for light nuclei, taking into account only a
  fixed number of
degrees of freedom, that
 i) automatically embeds  the whole successful phenomenology 
already developed within the non relativistic framework, and 
ii)  includes non perturbatively the relativistic
features requested  by the Poincar\'e covariance. 
The  Bakamjian-Thomas (BT) procedure (see, e.g., Ref. \cite{KP91}) allows us to  
implement the above program, since within such an approach it is possible to 
explicitly construct operators that fulfill the commutation rules of the
Poincar\'e generators, in terms of i) operators that do not contain the 
interaction 
and ii) 
an operator that acts as the mass of the interacting system. This interacting mass
operator, that depends upon intrinsic variables, has to satisfy 
 the same constraints  given by the Galilean group, namely
the same properties implemented  in the non relativistic quantum mechanics. In
summary, one can exploit 
 realistic wave functions for light nuclei
 (for A=3, see, e.g., Ref. \cite{KVR94}), that depend upon  suitable Jacobi
 coordinates,  in order to  evaluate 
 matrix elements of a 
Poincar\'e covariant current operator \cite{LPS98}  for an interacting 
system.   The appeal of the BT procedure, within the LF framework, is given by the fact that
 relativistic effects
imposed by Poincar\'e covariance can be straightforwardly taken into account
through the relativistic kinematics and 
the presence of the so-called Wigner-Melosh rotations (see, e.g., Ref. \cite{KP91}), 
that ultimately lead to using
 the standard Clebsh-Gordan machinery for  obtaining many-nucleon wave
functions with the correct angular coupling.
 
\section{Formalism} 
Among the  main features of   LF RHD (for an extended
review see, e.g., Ref. \cite{KP91}), one has to mention the
largest number of  kinematical Poincar\'e generators (seven, as dictated by 
 the symmetry of the
{\em  initial} hypersurface $x^+=0$) and the simplest procedure
for separating out the center of mass  motion from the intrinsic one, in strict
analogy with the non relativistic procedure (given the absence of the square root in
the operator  generating LF-time translation of the system). On one side, since the LF boosts
form a subgroup of the kinematical set, it is not necessary any Wigner function
for taking care of the boost transformation of the nuclear wave function,
 when we consider, e.g., a
nuclear final state recoiling in the laboratory frame. On the other side, since
the rotations around the $x$- and $y$-axes are dynamical ones, we need to
overcome a difficulty. A possible strategy is suggested by
the BT construction, that
amounts to put in relation
the LF spin with the canonical one  through unitary
transformations, namely Wigner-Melosh rotations.
For the sake of concreteness, the Wigner-Melosh rotations in the $2\times 2$
representation, are given by
\be
D^{{1 \over 2}} [{\cal R}_W (k^{\nu}_i)]_{\sigma\sigma'}=~
\chi^\dagger_{\sigma}~L_c^{-1}(k^{\nu}_i)~L_{LF}(k^{\nu}_i)
~\chi_{\sigma'}=\nonu =\chi^\dagger_{\sigma}~{m+k^+_i-\imath 
{\bf\sigma} \cdot (\hat z \times
{\bf k}_{i\perp}) \over \sqrt{\left ( m +k^+_i \right )^2 +|{\bf
k}_{i\perp}|^2}}
~\chi_{\sigma'}\ee
where $L_c^{-1}(k^{\nu}_i)$ and  $L_{LF}(k^{\nu}_i)$ are  
the ${\bf SL(2,C)}$ representations of  the canonical boost and the LF one,
respectively (see, e.g. Ref. \cite{LPS98}), and $k^{\nu}_i$ is the intrinsic 
momentum of  the
i-th constituent.

Therefore,  the canonical spin of a single constituent 
(i.e., the total angular momentum in its intrinsic
frame) is given in terms of  the LF one, by
\be
\vec s_c(i)=\left [{\cal R}_W (k^{\nu}_i) \right ]~\vec s_{LF}(i)
\ee
and  the component of   a state $|\psi \rangle$ in the
  LF-spin  basis is related to its 
  canonical counterpart, by
\be
_{LF}\langle \sigma|\psi\rangle =\sum_{\sigma'}
~D^{{1 \over 2}} [{\cal R}_W^\dagger (k^{\nu}_i)]_{\sigma\sigma'}~
_{c}\langle \sigma'|\psi\rangle
\ee
This illustrates the source of a very cumbersome algebra necessary to perform
Trinucleon calculations.

Following Ref. \cite{LPS98}, for an interacting system, an em 
 current operator, $J^\mu(0)$,
that fulfills  the extended Poincar\'e  covariance (i.e.
considering parity and time 
reversal, as well) and  
Hermiticity, can be constructed by  a suitable auxiliary operator,
$j^\mu$, that fulfills rotational covariance around the $z$-axis in 
a Breit frame (${\bf P}_f+{\bf P}_i=0$), where the $\perp$ component of the momentum transfer is vanishing
(${\bf q}_\perp=0$). Note that such a frame is  {\em different} from the Drell-Yan one, where
$q^+ = 0$, where the kinematical symmetry around $\widehat q$ is not exploited. In general,
 the matrix elements  $\langle P_f|J^\mu| P_i \rangle$, still acting on
 internal variables,   are directly given 
by the  matrix elements of the auxiliary operator $j^\mu$, evaluated in the
chosen Breit
frame. A possible Ansatz for  
a {\em many-body} auxiliary operator is built from i)
the free current (a one-body operator) and ii) the $\perp$ component of the 
angular momentum operator
$\vec S$ (a many-body operator in LF)
 as follows
 \be   
j^{\mu}_{fi}(q\hat{e}_z)={1 \over 2 }~\left [{\cal{J}}^{\mu}_{fi} (q\hat{e}_z) +
L^{\mu}_{\nu}[r_x(-\pi)]~e^{\imath \pi S_x}~{\cal{J}}^{\nu}_{if}(q\hat{e}_z)^*
~e^{-\imath \pi S_x} \right ] 
\label{curlf}\ee
where $r_x(\theta)$ is a rotation by an angle $\theta$ around the $x$-axis,
 $ 
{\cal{J}}^{\mu}_{fi}(q\hat{e}_z) = 
\Pi_f ~J_{free}^{\mu}(0) ~\Pi_i    
$, with
 $\Pi $ the projector onto the
  states 
of the  (initial or final) system and 
$\vec S \equiv$ the LF-spin operator of the system as whole: it acts on the
"internal" space.  The operator, $ J_{free}^{\mu}(0)$
 is the proper sum over A=2,3... free Nucleon current given by 
 \be J^\mu_{N}= -F^N_{2}(\Delta^2)~{(p^\mu+p^{\prime
\mu})\over 2M }+\gamma^\mu \left[F^N_{1}(\Delta^2)+F^N_{2}(\Delta^2)\right]\ee  where $\Delta^2=(p^{\prime
\mu}-p^\mu)^2$ and  $F^N_{1(2)}(\Delta^2)$  the Dirac (Pauli) Nucleon
ff. It should be pointed out that the   Nucleon ff depend upon
$ p^{\prime
\mu}-p^\mu$ and not upon $q^\mu\ne p^{\prime
\mu}-p^\mu$, since  only three components of the four-momentum,
i.e. $p^+$ and ${\bf p}_\perp$, are conserved quantities. On the other hand, 
within RHD framework all the particles are on their mass-shell.

In the chosen    Breit frame,  charge normalization and current conservation 
(for $M_f=M_i$) can be fulfilled by imposing 
 $ {\cal{J}}^{-}(q\hat{e}_z)={\cal{J}}^{+}(q\hat{e}_z)$\cite{LPS98,LPS2K}. 
  
Summarizing, if  $J^\mu(0)$, that is an operator acting in the whole space,
is Poincar\'e  covariant, then the intrinsic operator,
$j^{\mu}$,  is 
invariant for rotations around the $z$-axis, and viceversa. Moreover, in Eq.
(\ref{curlf}),  $S_x$ generates many-body contributions to $j^{\mu}$,  as well
as 
${\cal{J}}^{\mu}_{fi}(q\hat{e}_z)$, if   a many-body term is added to $J_{free}^{\mu}(0)$, as
discussed in the following Section.

For evaluating matrix elements of $j^\mu(q\hat{e}_z)$, the eigenstates of the 
 interacting system are needed. To this end, one can use the "non relativistic
solutions", but with Wigner-Melosh rotations in the angular part, if the interaction
$V\equiv M_{int} -M_0$ (where $M_{int} $ is the mass operator of the interacting 
 system and $M_0$ the corresponding free mass) can be embedded in a BT
framework.
The BT construction for obtaining interacting Poincar\'e generators suggests 
a necessary (not sufficient) 
condition  on the interaction (see Ref. \cite{KP91}): $V$
 must depend upon intrinsic variables  combined
in scalar products, i.e.
$ [\vec{\cal B}_{LF}, V ] = [\vec S_{0}, V ] = 
[P_{\perp}, V ] = [P^+, V ]= 0
$,
where $\vec{\cal B}_{LF}$ are the LF boosts, $\vec S_{0}\equiv$  the
angular momentum operator for the non interacting case (since
$S^2_{0}=S^2_{int}$ and $S_{0,z}=S_{int,z}$, the eigenvalues of $S^2_{0}$
and $S_{0,z}$ can label the eigenstates of the interacting system). The non
relativistic interaction fulfills the above requirements.

 \section{Trinucleon em observables}

\begin{figure}[t]
\label{2bcur}
\includegraphics[width=2.8 cm,angle=-90] {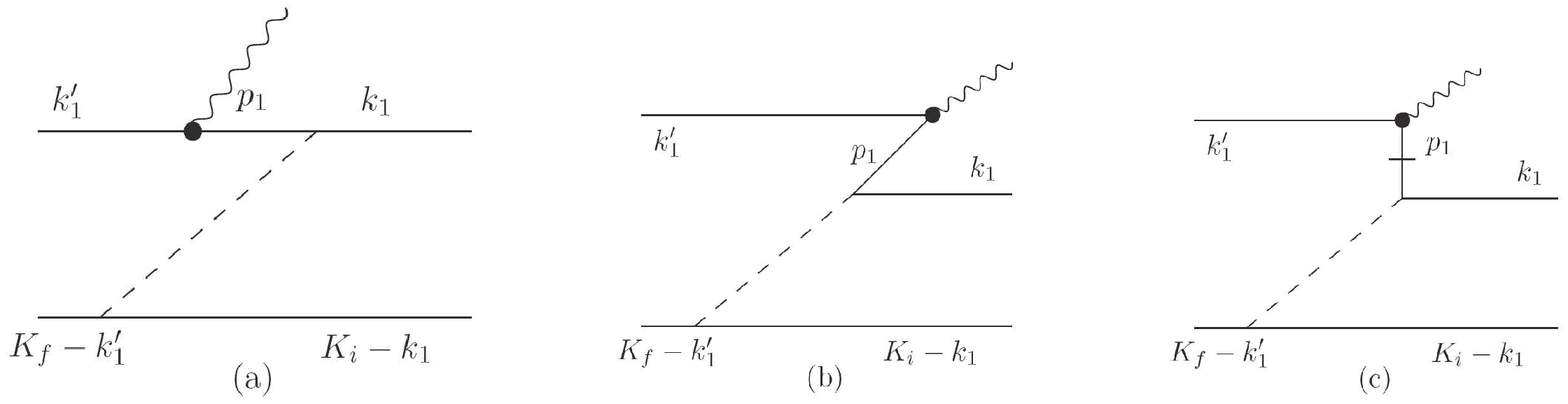}
\includegraphics[width=2.8 cm,angle=-90] {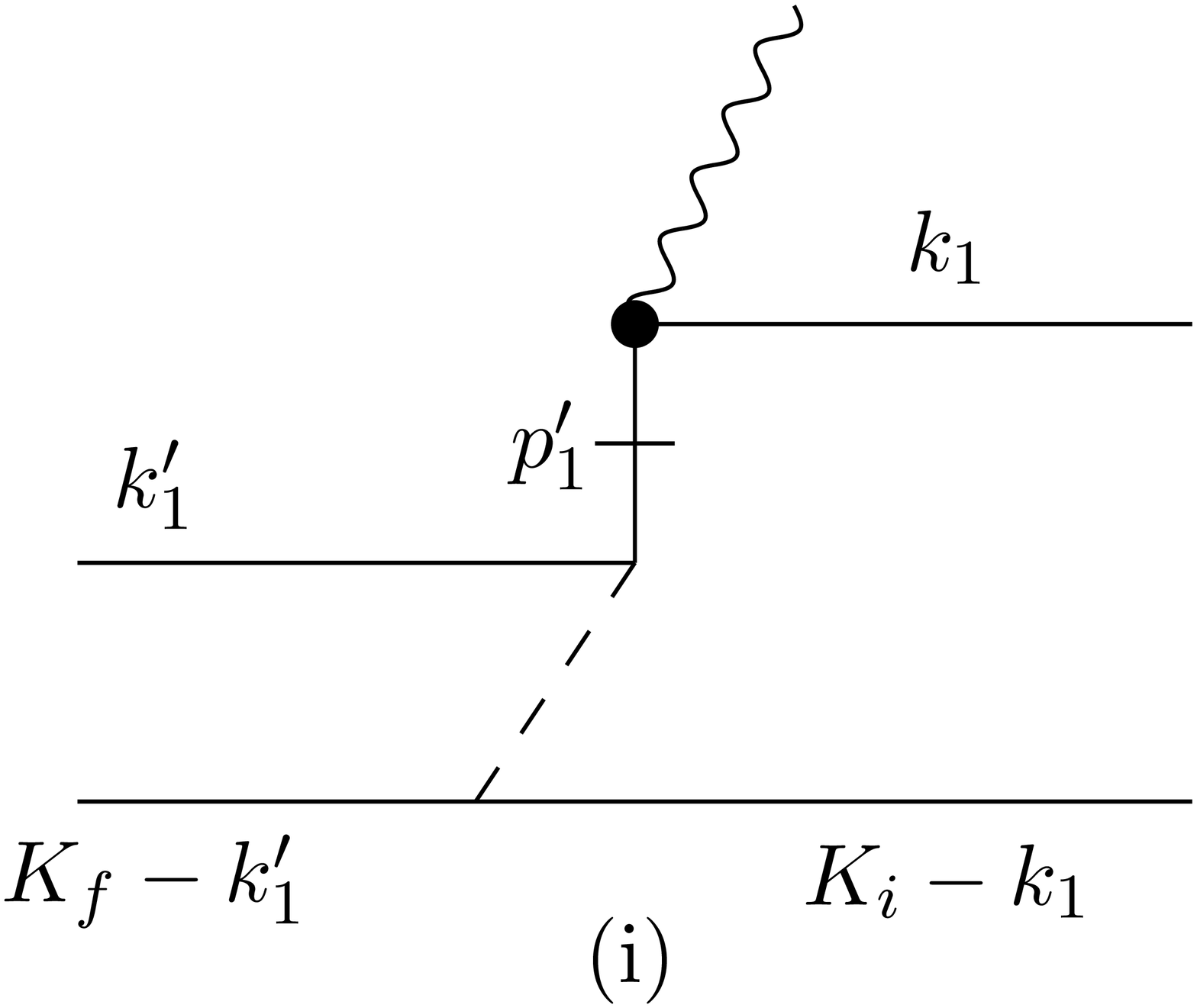}
\caption{A class of two-body (at least three particles
in flight) current contributions generated 
 within a LF analysis of a Yukawa model, see
text. Diagram (b) represents a pair production contribution. Diagrams (c) and  (i)
contain the 
instantaneous, in LF time, propagation of an internal fermionic 
particle,  marked by
a horizontal dash.  The LF time flows from the right to the left. (After
\cite{adnei08})}
\end{figure}

The choice of a   Breit frame where ${\bf q}_\perp=0$, namely  
$q^+ \ne 0$, is a far reaching one, since, as above mentioned,
one can find a simple constraint to be fulfilled by a one-body intrinsic 
current
operator for recovering the Poincar\'e covariance, as well as by each many-body
 term that one could add to $J^\mu_{free}$. Furthermore, this
 choice  necessarily produces  a two-body
current related to a pair production \cite{adnei08} (see diagram (b) in
Fig. \ref{2bcur}), compelling us to consider the inclusion of a larger set of
two-body currents  in the future
calculations. In particular, a recent   analysis of  a 
4D  Yukawa model \cite{adnei08}, in ladder approximation, has led to
 a 3D current on the LF,  fulfilling the Ward-Takashi Identity
  (for a general discussion see \cite{adnei08}). In Fig. (\ref{2bcur}),
 a set of contributions to the first-order (in the interaction) 3D
LF current is shown. In the preliminary 
calculations of the Trinucleon em observables
presented in this contribution, the  two-body terms, like the ones depicted 
in Fig. (\ref{2bcur}),
are not included, while the application to the Deuteron 
case is in progress. One can
anticipate that i) the pair term  affects all the three Deuteron ff,
while  instantaneous terms  (present  only for  fermionic constituents) 
contribute to the magnetic one, ii) the pair term vanishes for $q^+ \to
0$, as it must do, while  instantaneous ones survive, iii) 
the remaining,  on-mass shell terms (like (a)  in Fig. (\ref{2bcur}))
 affect all the  ff in the whole range of $q^+$, iv) the pair term
should be maximal at $q^+\sim m_N$ (cf the discussion in \cite{MFPS02}). 
The macroscopic current of the Trinucleon is given in terms of charge and
magnetic ff by  
\be
J^\mu_{T_z}= -F^{T_z}_{2}(Q^2){(P^\mu+P^{\prime \mu})\over 2M_{T_z}} 
+\gamma^\mu (F^{T_z}_{1}(Q^2)+F^{T_z}_{2}(Q^2))\ee 
where $T_z=\pm 1/2$ labels $^3$He and $^3$H respectively. 
Microscopic evaluations of the ff are based on proper traces 
of the current in  Eq.
(\ref{curlf}), viz \be
    F_{ch}^{ T_z}(Q^2) = F^{T_z}_{1}-{Q^2 \over 4~M^2_{T_z} } F^{T_z}_{2}=
    {1 \over 2} ~ Tr[j^+(T_z)]= {1 \over 2} ~ Tr[{\cal{I}}^+(T_z)]
   \nonu  F_{mag}^{ T_z}(Q^2)= F^{T_z}_{1}+F^{T_z}_{2}
   = - i { M_{T_z} \over Q} ~ Tr[\hat{\sigma}_y ~
    j_x(T_z)] = - i { M_{T_z} \over Q} ~ Tr[\hat{\sigma}_y ~
    {\cal{I}}_x(T_z)]   
\ee
where the matrix elements of the microscopic current are given by 
\be     {\cal{I}}^{r}_{\sigma' \sigma }(T_z) \equiv 
    \langle \Psi_{{1 \over 2} \sigma' ~}^{{1 \over 2} T_z},
    {P}' | ~ {\cal J}^{r} ~ |\Psi_{{1 \over 2} \sigma ~
   }^{{1 \over 2} T_z}, {P} \rangle 
 \ee
 with $|\Psi_{{1 \over 2} \sigma}\rangle$  
the Trinucleon wave function. In the actual calculation we have  used the 
 bound states obtained  through a variational technique  \cite{KVR94}
 by taking into account  two-nucleon forces, like   
 $AV18$ \cite{AV18}, and three-body
ones, like $UIX$ \cite{UIX}. Moreover, $^3$He and $^3$H are distinct, 
since the Coulomb forces are
included. The bound states contain 
 $S$, $S^\prime$, $P $ and $D$ waves, and 
the Wigner-Melosh rotations suitable for expressing the LF spins in terms of the
standard ones. It should be pointed out that the numerical calculations involve
 6D Montecarlo integrations. 

In Tables \ref{tab1} and \ref{tab2}, preliminary results for the static em
properties of $^3$He and  $^3$H are shown. In particular in the first Table,
 the
bound states obtained by retaining only two-nucleon forces are presented, while
in the second Table the bound states correspond  to $AV18 + UIX$. 
The benefits of considering the Poincar\'e covariance are
clear and of the same order found in the Deuteron case \cite{LPS2K}. We expect
a possible improvement of the magnetic moments by including the instantaneous
contributions (cf Fig. \ref{2bcur}), since 
they particularly affect the magnetic ff only (as already seen for the Nucleon
case \cite{MFPSS}).

\begin{table}[htb]
\label{tab1}
\caption {Preliminary calculations of magnetic moments and charge radii of 
$^3$He and  $^3$H.
 The two-body force, $AV18$, and the Coulomb interaction are  included.
  Trinucleon wave functions from \cite{KVR94}. Probability of the waves
  considered:
${\cal P}_{S+S'}(Av18)\sim ~91.4$\%,    
${\cal P}_{P}(Av18)\sim~0.07  $\%, ${\cal P}_{D}(Av18)\sim~8.5  $\% }
 
\begin{tabular} {|c||c| c|c| c|}
\hline
Theory &  $\mu(^3\rm{He})$ & $\mu(^3\rm{H})$ & $r_{ch}(^3\rm{He})$fm  &$r_{ch}(^3\rm{H})$fm \\
\hline
NR(S+S') & -1.700(1)  & 2.515(3)& 1.926(3) &1.726(3)  \\
LF(S+S')  & -1.758(1)  & 2.600(3) & 1.949(3) &1.771(3)  \\
NR(S+S'+P+D) & -1.762(1) & 2.579(2) & 1.916(4)   & 1.718(4)  \\
LF(S+S'+P+D)  & -1.834(2)& 2.674(2) & 1.941(4)  & 1.759(4) \\
\hline
Exp. &  -2.1276 & 2.9789 & 1.959(30) & 1.755(86) \\
\hline
\end{tabular}
\end{table}

\begin{figure}
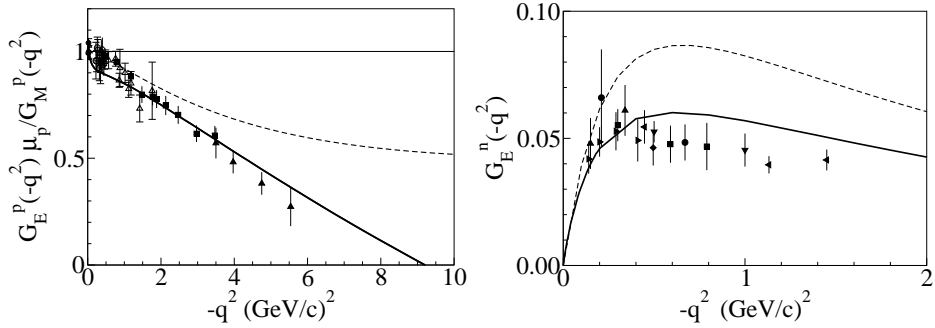

\includegraphics[width=6.1cm,angle=0] {FF_GK_ratio_Tr.eps}
\includegraphics[width=6.1cm,angle=0] {FF_GK_Gen_Tr.eps}
\caption{Left panel: $\mu^p G_E^p(Q^2)/G_M^p(Q^2)$. Right panel: charge
neutron ff.
Solid line: LF Nucleon ff \cite{MFPSS}.  Dashed line: Gari-Kr\"umpelmann
\cite{GK}}
\label{chn}\end{figure}

\begin{figure}
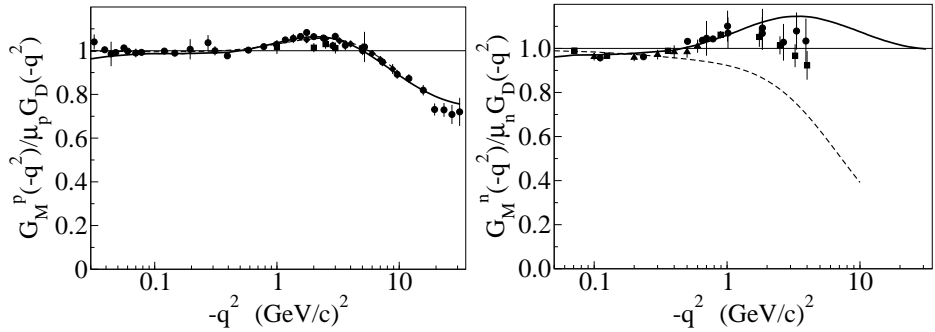

\includegraphics[width=6.1cm,angle=0] {FF_GK_Gmp_Tr.eps}
\includegraphics[width=6.1cm,angle=0] {FF_GK_Gmn_Tr.eps}
\caption{The same as in Fig. \ref{chn}, but for the magnetic Nucleon ff.}
\label{mgn}
\end{figure}

\begin{table}[htb]
\label{tab2}
\caption {Preliminary calculations of magnetic moments and charge radii of
 $^3$He and  $^3$H.
 Two- and three-body forces, $AV18+UIX$, are included, as well as the Coulomb
 interaction. 
Trinucleon wave functions from \cite{KVR94}. Probability of the waves
  considered:
${\cal P}_{S+S'}(Av18+UIX)\sim ~90.5$\%   
${\cal P}_{P}(Av18+UIX)\sim~0.01  $\% ${\cal P}_{D}(Av18+UIX)\sim~9.3  
$\%.}

\begin{tabular} {|c||c| c|c| c|}
\hline
Theory &  $\mu(^3\rm{He})$ & $\mu(^3\rm{H})$ & $r_{ch}(^3\rm{He})$fm  &$r_{ch}(^3\rm{H})$fm \\
\hline
NR(S+S') & -1.697(1)  & 2.494(2)& 1.848(3)   &1.695(3)   \\
LF (S+S')  & -1.759(2)  & 2.588(2) & 1.870(3)   &1.712(3)   \\
NR(S+S'+P+D) & -1.760(1) & 2.569(2) & 1.841(4)    & 
1.666(4)   \\
LF (S+S'+P+D)  & -1.837(2)& 2.669(2) & 1.867(4)   & 
1.690(4) \\
\hline
Exp. &  -2.1276 & 2.9789 & 1.959(30)  & 1.755(86)  \\
\hline
\end{tabular}
\end{table}
In the evaluation of the Trinucleon ff two different sets of Nucleon em ff  have been
considered: i) the Gari-Kr\"umpelmann Nucleon ff \cite{GK} and ii) the ones
obtained within a novel LF approach \cite{MFPSS}, see Figs. \ref{chn} and \ref{mgn}, in order to test the
dependence of the Trinucleon ff upon the new features of the Nucleon ff, like
a possible zero in the proton charge ff \cite{MFPSS,JLAB}. 
 
\begin{figure}
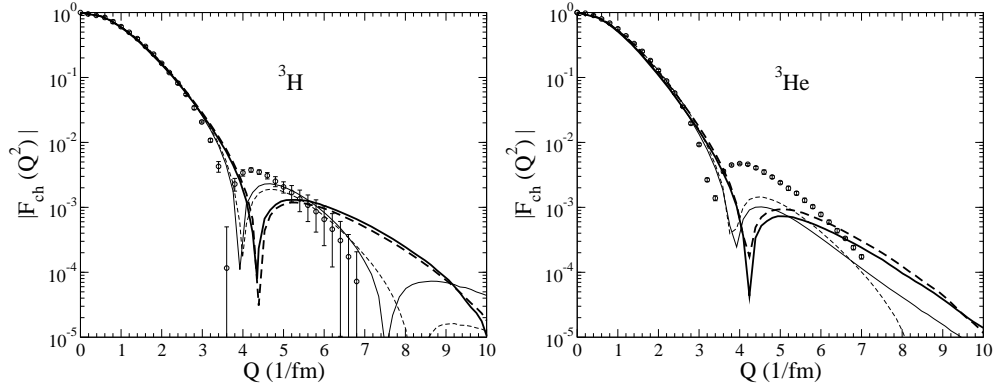

\includegraphics[width=6.5cm,angle=0] {Fch_H3n_Tr.eps}
\includegraphics[width=6.5cm,angle=0] {Fch_He3n_Tr.eps}
\caption{Preliminary calculations of $^3$H and $^3$He charge ff
  vs $Q^2$. The adopted
Trinucleon wave functions \cite{KVR94} contain both
the  $AV18$ two-nucleon interaction and the Coulomb interaction. 
$S+S^\prime+P+D$ waves are taken into account.
Thick lines: LF calculations.  Solid line: 
 full calculation  and   LF
Nucleon ff \cite{MFPSS};
dashed line:  full calculation and Gari-Kr\"umpelmann Nucleon ff \cite{GK}.
Thin lines: non relativistic calculations. Data from \cite{Sick}.}
\label{fch2b}\end{figure}

\begin{figure}
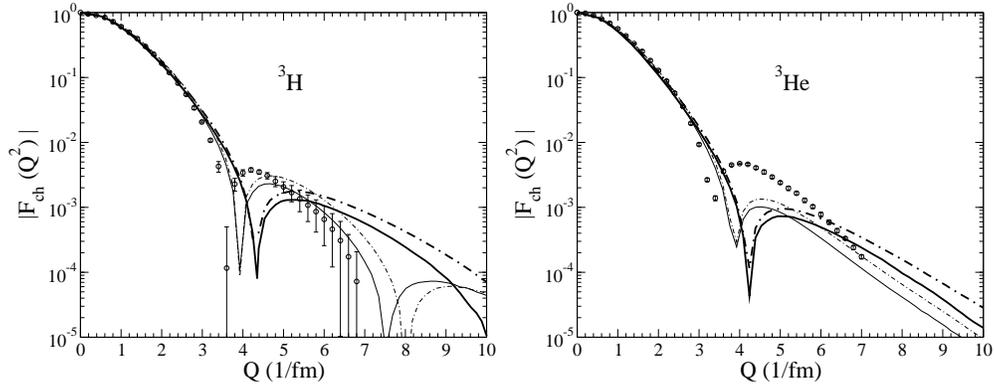

\includegraphics[width=6.5cm,angle=0] {Fch_H3_3BF_Tr.eps}
\includegraphics[width=6.5cm,angle=0] {Fch_He3n_3BF_Tr.eps} 
\caption{Preliminary calculations of $^3$H and $^3$He charge ff  vs $Q^2$. The adopted
Trinucleon wave functions \cite{KVR94} contain both i) two-nucleon and three-body forces:
 AV18 +UIX and ii) Coulomb interaction. 
$S+S^\prime+P+D$ waves are taken into account. Thick lines: LF calculations. 
Solid line: 
full  calculation with the $AV18$ two-nucleon interaction    \& LF Nucleon ff \cite{MFPSS}.
Dash-dotted line:   calculation with two-nucleon+three-body interactions 
($AV18+UIX$) \& 
 LF Nucleon ff. Thin lines: non relativistic calculations. Data from \cite{Sick}.}
\label{fch3b}
\end{figure}

\begin{figure}
\includegraphics[width=6.5cm,angle=0] {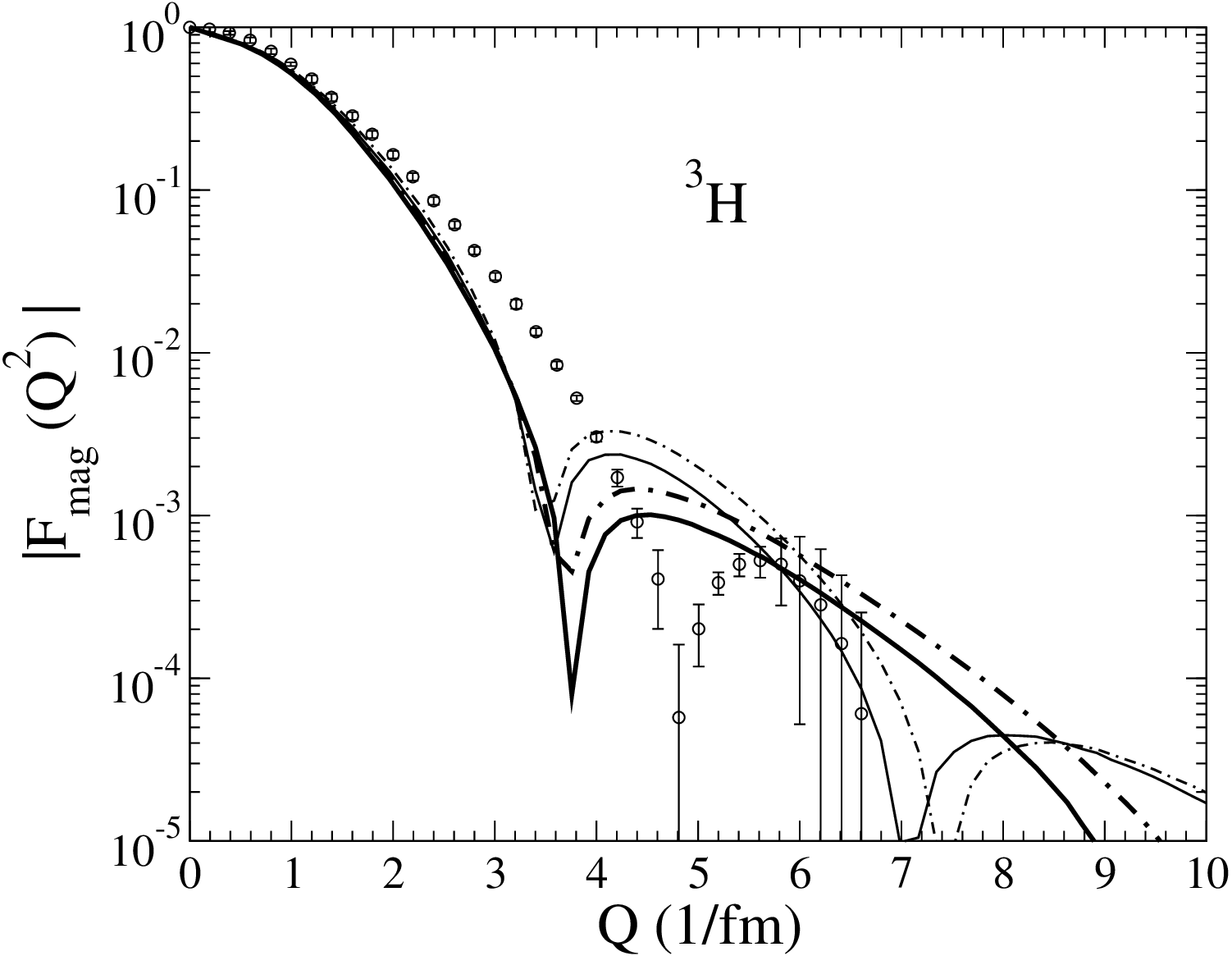}
\includegraphics[width=6.5cm,angle=0] {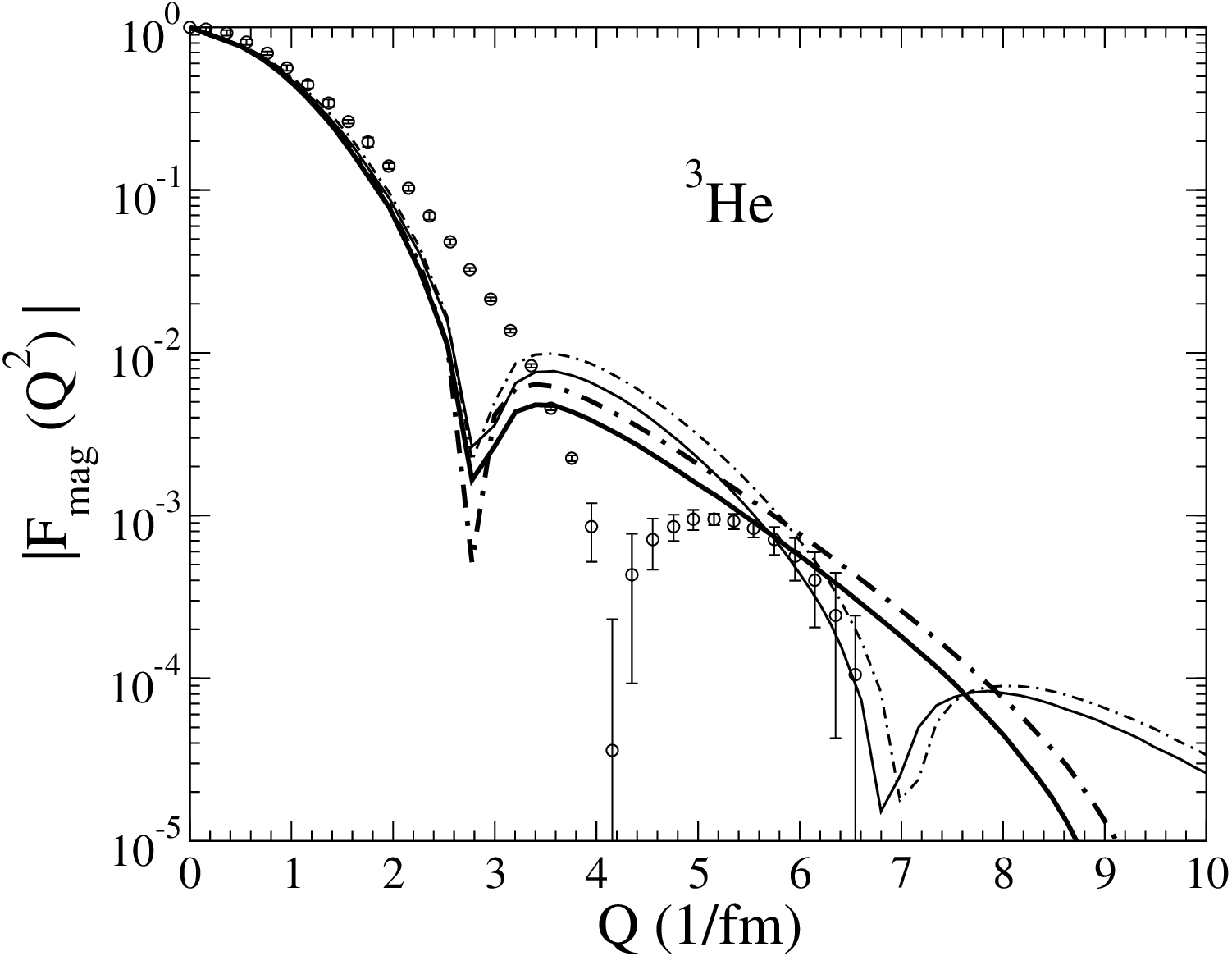} 
\caption{The same as in Fig. \ref{fch3b}, but for the magnetic ff.}
\label{fmg3b}
\end{figure}

The charge and magnetic ff of $^3$H and $^3$He, evaluated in the  Breit
frame where ${\bf q}_\perp=0$ and with $S$, $S^\prime$, $P$ and $D$ waves, are shown in Figs. \ref{fch2b} and \ref{fmg2b}
for bound states corresponding to two-nucleon forces, while in  Figs. \ref{fch3b} and \ref{fmg3b}
the calculations with three-body forces are  presented. Calculations with only
$S+S^\prime$ waves are shown in \cite{BPS07}. Some expected
features, like the presence of large relativistic effects on the tails for $Q
\geq 7~(1/fm)$, are well
confirmed, as well as the necessity of two-body dynamical corrections to the current
operator (cf the positions of the minima). An interesting signature of 
the three-body forces can be found in the tails, since they give more
 binding and 
smaller charge radii. It could be relevant, if such an a effect
 will  still be  present 
 after including two-body dynamical currents.
As a final remark, we should note that
differences
between relativistic calculations  obtained by using  the  Gari-Kr\"umpelmann ff 
and 
the ones calculated adopting the  LF ff  
are not sizable.
\begin{figure}
\includegraphics[width=6.5cm,angle=0] {Fmag_H3n_Tr.eps}
\includegraphics[width=6.5cm,angle=0] {Fmag_He3n_Tr.eps}
 \caption{The same as in Fig. \ref{fch2b}, but for the magnetic ff.}
\label{fmg2b}\end{figure}

\section{Conclusions \& Perspectives}
 In order to embed the Poincar\'e covariance in the description of light nuclei
 we 
 adopt  a
 Light-Front RHD and  the Bakamjian-Thomas procedure. Extending our approach,
 already applied to the Deuteron case \cite{LPS2K}, 
 the em observables of $^3$He and $^3$H
  have been calculated for the first time with all the waves, $S$, $S^\prime$, $P$ and $D$,
 in the bound states and taking into account the three-body forces as well.
 The  relativistic effects on em observables at $Q^2=0$, though of the order of few \% but 
 in the correct direction, are  encouraging. Moreover,
 the sizable effects at high $Q^2$ indicate the essential role played by 
 the Poincar\'e covariance
 for analyzing the em ff in the region of few GeV's. Notably,  
  three-body forces could be important in the same kinematical region.
 
 A full calculation, with a systematic inclusion of two-body currents, like the
 ones shown in Fig. \ref{2bcur},   will be presented
  elsewhere.

\end{document}